*Article*

# Between Post-Flâneur and Smartphone Zombie: Smartphone Users' Altering Visual Attention and Walking Behavior in Public Space

**Gorsev Argin [1,*], Burak Pak [2] and Handan Turkoglu [1]**

[1] Department of Urban and Regional Planning, Istanbul Technical University, Istanbul 34367, Turkey; turkoglu@itu.edu.tr
[2] Department of Architecture, KU Leuven Campus Sint-Lucas Brussels and Ghent, 9000 Ghent, Belgium; burak.pak@kuleuven.be
* Correspondence: arging@itu.edu.tr

**Abstract:** The extensive use of smartphones in our everyday lives has created new modes of appropriation and behavior in public spaces. Recognition of these are essential for urban design and planning practices which help us to improve the relationship between humans, technologies, and urban environment. This study aims to research smartphone users in public space by observing their altering visual attention and walking behavior, and, in this way, to reveal the emergent "new figures". For this purpose, Korenmarkt square in Ghent, Belgium, was observed for seven days in 10-min time intervals. The gaze and walking behavior of smartphone users were encoded as geo-located and temporal data, analyzed and mapped using statistical and spatial analysis methods. Developing and implementing new methods for identifying the characteristics of smartphone users, this study resulted in a nuanced characterization of novel spatial appropriations. The findings led to a better understanding and knowledge of the different behavior patterns of emergent figures such as "post-flâneurs" and "smartphone zombies" while uncovering their altering visual interactions with and movements in the public space. The results evoked questions on how researchers and designers can make use of spatial analysis methods and rethink the public space of the future as a hybrid construct integrating the virtual and the physical.

## 1. Introduction

The rapidly altering nature of public place and its significance in our everyday lives has been one of the most compelling queries in urban studies, particularly beginning with the dominance of the private over public life [1]. Since the early 1990s, Information and Communication Technologies (ICTs) have become one of the main actors of this alteration, especially when the Internet introduced us to the virtual space—a new means of social interaction—that we can wander through from our personal–private spaces. With the emergence of new forms of these "third places" [2], the virtual digital space became a public sphere, thus igniting discussions about the decline of the physical public space [3]. However, with the introduction of mobile modes of ICTs (mICTs), people began carrying this virtual space along with them into the public space and leading to the emergence of "hybrid space" [4–7], which has been used to define a variety of multi-layered constructions of space in the literature. When it comes to the mICTs, hybrid space is used to describe mainly two kinds of juxtapositions of dualities: private–public [4–6] and virtual–physical [7]. These hybrid spaces are not static and continuously reconstructed in public spaces by the people who carry mICTs [7]. In this article—the preliminary and short version of which was presented at a conference [8]—we define hybrid space as a hybrid construct composed of the physical and the virtual space. We suggest an understanding of "hybrid public space" combining human and non-human, technological and non-technological actants, including mobile and non-mobile devices and the Internet of Things (IoT). We know that today, new spontaneous and non-traditional appropriations and behaviors in public spaces emerge as people interact with mobile technologies, thus leading to a gradual transformation of public spaces so as to accommodate these new behaviors [9] and raising a query about the hybrid public space of tomorrow.

Today, smartphones—one of the most widely used interfaces in hybrid space—prove to be one of the most rapidly adopted technologies in history, and they have already become an essential part of our everyday lives [10]. Ubiquitous technologies such as smart glasses, smartwatches, and even smart contact lenses render mICTs to be more invisible and embedded in the human body. Therefore, examining the effects of smartphones on the appropriation of public spaces is essential while they are still visible as an external mICT and observable through the "new" figures of public space.

Our literature study uncovered references to two main "new" walking figures of public space whose visual attention is divided between the virtual and the physical space: (1) "smartphone zombies" who walk slowly as a result of their great attention to the virtual through the smartphone screen, and (2) "post-flâneurs" who shuttle between the virtual and the physical space with a pursuit of urban experience in a similar fashion to their pioneers—*flâneurs*—who have been used strolling as a means of experiencing the city instead of arriving at a certain destination [11]. These two figures were mentioned in a few academic studies; however, they were not observed together in an empirical study.

A significant amount of the available literature on smartphone use whilst walking defines it as a multitasking activity and addresses the topic from the safety perspective by defining smartphone users as "distracted walkers" [12–19]. The contributions offered by this perspective hardly go beyond solutions to avoid distraction and regulate pedestrian behavior, such as minor physical interventions to the public space and interface-based solutions. Current physical design interventions, like pavement lights at pedestrian crossings, can serve this purpose to some extent. Furthermore, smartphone interface-based interventions have been proven to fail in reducing the risks related to smartphone distractions, since they draw more attention away from the environment as opposed to what they were originally intended for [20]. None of the reviewed studies on the behavior of smartphone users involved naturalistic observations in the form of geo-information, thus failing to register the interactions of the users with the public space at the architectural scale.

In this sense, a better and nuanced understanding of the altered visual attention and walking behavior beyond the safety risks, as well as its effects on the appropriation of public spaces, is essential both to understand the public spaces of today and to define and design the "hybrid public spaces" of tomorrow. With these in mind, our study was designed as a naturalistic observation that aims to understand the altering visual engagement of smartphone users with their surroundings while they walk through a public space by examining the movements of their gaze and body. By examining the gaze of the smartphone users, we aimed to understand how visual attention was divided between the virtual and the physical space, and by converting these observations into geo-located information, we gained a better insight into how walking behavior changes depending on the altering object of visual attention. Based on these statistical and spatial analyses, we identified new figures among smartphone users in the public space.

By analyzing the temporal configuration of gaze shifts and the spatial organization of their movement with an observational survey, this study addresses the following research questions:

- RQ1. Can we identify different figures among smartphone users in public space by observing their gaze direction and movement in the public space?
- RQ2. How does smartphone users' visual attention in public space change depending on different modes of smartphone activities?

- RQ3. How do smartphone users' walking speed and movement in the public space change depending on smartphone activities?
- RQ4. What are the implications of these figures and their movement in the public space in terms of hybrid public space and its design?

To address these research questions, we will present an in-depth literature review on pedestrians with smartphones, the emerging figures, and interventions (Section 2). The empirical study will follow this section, where a public square in Ghent, Belgium, was observed for seven days in 10-min time intervals between 20 October and 14 December 2018, and 350 smartphone users (198 female, 152 male) among the passers-by were examined (Sections 3 and 4). Finally, we will discuss our findings and present conclusions on the emergent figures among smartphone users in public spaces and the implications of the results for the hybrid public space (Sections 5 and 6).

## 2. Pedestrians with Smartphones in Public Spaces: The Emergent Figures and Interventions

As of 2017, over 65% of the world's population owned mobile devices, and over half of them were smartphone users [21]. Countries with advanced economies take the lead, with a median rate of 76% of smartphone ownership, and these rates reach even higher among young adults [22]. This rapid adoption of smartphones in our daily lives made them both the instrument and the subject of an increasing number of academic studies. Most of these studies focus on the substantial impact of smartphones on walking [12,13,15–20,23–33]. The observed impact on gait indicates higher cognitive loads and lower attention to the surrounding environment [29]. Studies that associate walking and using a smartphone mainly focus on the cognitive and behavioral effects of this multitasking action and examine the adverse effects of smartphone distraction on attention and walking behavior [12,13,17,19,23,24,27,29].

The majority of the research on smartphone use in public spaces focuses on its effects on attention, and particularly on visual attention. This is mainly due to the fact that smartphones, with their ever-enlarging screen sizes and increasing conduciveness, attract the visual attention of users even more than their pioneers, cell phones. The shuffling gaze between the smartphone screen and the surrounding environment also shows its visible effects on walking behavior, which proves to be the second most studied topic in the available literature. For instance, studies demonstrate that walkers distracted by smartphones tend to have a slower gait speed [12,14,32,34]. Although an overwhelming amount of the related literature is focused on these two effects, smartphone activities that create the distraction differ from one study to another.

Different smartphone activities create various distractions while walking, and these can be grouped as cognitive, visual, audio, and physical distractions [35]. Some of these activities lead to more than one type of distraction, and each has effects on attention and gait to different extents. For instance, speaking (talking) on the phone creates a cognitive distraction, while typing (texting) and reading require repeatedly diverting eyes away from the surrounding environment [33], thus causing cognitive and visual distractions [36]. Listening to music, on the other hand, is an auditory distraction; thus, it is less unfavorable for walkers than the other kinds of distractions [36,37]. Holding the mobile phone without using it, which has been investigated by a limited number of studies [30,38], appears to be a less-destructive type of mobile device usage. However, its significant prevalence [30,38] calls for further research.

Other activities that became more and more smartphone-dependent over the last two decades are photo-taking and navigating. Photo-taking (also video-recording) has not yet been studied in terms of smartphone use while walking. However, estimates indicate that, in 2017, approximately 85% of digital photos were captured with a smartphone [39]. Thus, the increasing use of smartphones for photography and video-recording as well as joint activities—such as photo/video editing and sharing applications—highlights the necessity of further analysis of their effects on attention and walking behavior. Navigation is another everyday smartphone activity that directly affects the gait characteristics. Laurier et al. [15] found that map actions, such as displaying the map or monitoring the scale, correlate closely with the walking actions such as unilateral-stopping, turning, and restarting.

### 2.1. Observing Pedestrians with Smartphones in Public Spaces

A significant number of studies on smartphone usage in public spaces are based on self-report data [25,36,40,41]. However, researches indicate that self-reported usage of smartphones hardly reflects the actual use [42,43]. A research study [42] showed that the actual use amounted to more than double the figures reported by users themselves. Therefore, observational research is essential to understand the actual usage rates and the effects of smartphone use in public spaces.

A significant amount of the observation-based studies on this phenomenon addresses the topic from a pedestrian safety perspective. In this context, many of the related studies observed the walking behavior of smartphone users in a simulated environment in which they can restrict environmental factors [13,16,17,20], and most of the real-world experiments were conducted at street crossings [12,14,18,31,33]. Although crossings have lower smartphone usage rates compared to other public spaces, they comprise most of the literature given that they pose the highest risk in terms of pedestrian safety [36].

There is still a relatively small number of studies on smartphone use in public spaces. These observation-based studies examine the changing walking behavior of smartphone users in real-world public spaces based on egocentric [15] or pre-scripted walking experiences [19].

There is a limited number of naturalistic observation studies that examine smartphone usage in public spaces. A study from 2010 on cell phone users in public spaces focused on talking on a mobile phone while walking, and it showed that mobile phone usage can cause ”inattentional blindness” [37]. Another study examined walkers who carry mobile phones without using—whom they called “phone-walkers”—and suggested illustrative patterns through users’ demographic characteristics [30]. A passive-observation study conducted by Hampton and colleagues [44] replicated William Whyte’s pioneer study titled *The Social Life of Small Urban Spaces* [45] after 30 years by recording videos of the same public spaces between 2008–2010. The study revealed that the percentage of people who use a mobile phone in public spaces is not significant, and people tend more to use a smartphone when they are walking alone compared to when they are in a group. They also found that mobile phone users are more likely to linger in public spaces, which underlines the fact that technology is not disconnecting society [44], as some theorists assert. However, given the boost in mobile phone usage in everyday life as a consequence of the rapid spread of the smartphone technology, there is no room for doubt about the necessity for further research on the topic. A recent observational study [38] conducted in Spain also confirms this since they found that almost one-third of the pedestrians interacted with the smartphone while walking, and more than one of each ten pedestrians were behaving as a smartphone zombie. Their study also emphasizes the influence of gender, age, and city area on pedestrians’ smartphone usage.

Another recent observational study, which compares the walking speed of smartphone users and non-users in different public spaces, revealed that smartphone users are slower than non-users, and they also slow down their followers in narrower sidewalks. However, the researchers did not find any statistically significant difference between those walking on a pedestrianized bridge and deduced that smartphone use is more likely to affect walking speed depending on the intensity of pedestrian traffic, demographics, and the time of day [32].

### 2.2. New Figures in Public Spaces

This part of our study provides a review of the existing literature on two relatively new figures in public spaces: post-flâneurs and smartphone zombies. Located at the opposite ends of the continuum of smartphone users in public spaces, post-flâneurs walk and experience the city by traveling between the virtual and the physical space by means of smartphones, while smartphone zombies focus mostly on the virtual space through their smartphones as they walk in the physical space (Figure 1).

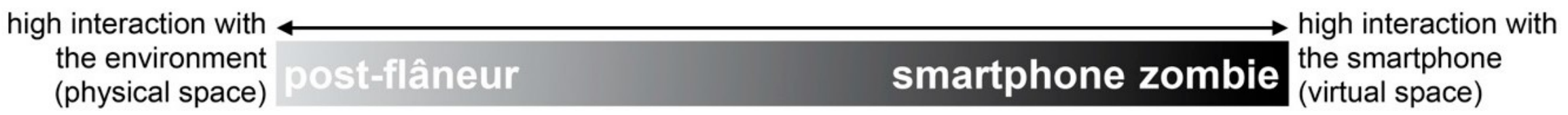

**Figure 1.** Continuum that shows smartphone users in public space based on the object of their visual attention with post-flâneurs and smartphone zombies at opposite ends (original figure by the authors).

The modern city of the 19th century introduced the concepts of flâneur and flânerie. Benjamin's [11] flâneur, a 19th-century character firstly portrayed by Baudelaire, strolls through the city and employs walking as a way of experiencing the city. A flâneur does not have the purpose of getting somewhere; the act of walking is the purpose itself. The fundamental requirements of a flâneur are the sense of presence, attention on both the urban environment and the passers-by, and a particular and slower pace making a place and experiences memorable [46]. Although flâneurs walk slowly, their eyes and minds wander swiftly [47]. Today, advances in mobile technologies created a new urban wanderer, which can be called the "post-flâneur". Post-flâneurs wander their gaze around the city and use smartphones and location-based applications that come with as a means of urban experience, such as taking photographs and sharing them on social media [48]. Furthermore, they re-appropriate the public space by "strolling, recording, and sharing it" [49].

Besides the post-flâneurs, new figures among smartphone users can be detected by examining the same variables mentioned before. Today, people who walk as they look at their smartphones are categorized as a new character that is more and more prevalent in public spaces. These new characters are referred to as "smartphone zombies" (smombies or head-down tribe) not only in popular culture but also in academic studies [28,38,40,50]. As the name indicates, they "walk slowly and without paying attention to their surroundings because they are focused upon their smartphone" [26]; thus, they are mainly studied from the perspective of pedestrian safety and technology addiction.

The main difference between these two figures is that post-flâneurs are assumed to split their attention between virtual and physical, while smartphone zombies highly concentrate on the virtual while presenting and walking in the physical space. The "attention" that we talked about here is a multidimensional concept, indeed. However, due to this naturalistic-observation study's limitations, we focus only on visual attention, which is detectable through the direction of smartphone users' gazes. Smartphone users switch gazes between the smartphone screen and the urban environment whilst walking, thus affecting their visual attention and walking behavior. This study will try to detect post-flâneurs and smartphone zombies in public space by examining their altering visual interactions with their surroundings and bring a new understanding of the concepts based on their altering walking behavior and spatial appropriation of the public space.

### *2.3. Existing Public Space Design Interventions Addressing Smartphone Users*

Today, public space design interventions that target smartphone users can also be examined in a spectrum ranging from smartphone zombies to post-flâneurs (Figure 2). Most of the existing interventions target smartphone zombies and try to ensure their own and others' safety with the help of some additions to existing public spaces. The main aim of these interventions is drawing attention back to the physical environment. They appear as traffic signs to warn vehicles and raise the awareness of pedestrians, some technologies such as radars and thermal cameras that give insights to smartphone applications designed to warn smartphone users [51], and additions to sidewalk pavements in the form of lights [52] and writings [53]. Such interventions were implemented mainly on/near pedestrian crossings where the gaze on a smartphone screen can result in vital pedestrian safety issues. In addition to these, there are some installations such as phone lanes [54] that have been drawn in dense pedestrian areas to protect non-users from the slowness and inattention of smartphone zombies by providing designated lanes where smartphone zombies can have indiscrete attention on the virtual. These interventions have also been used as awareness-raising campaigns.

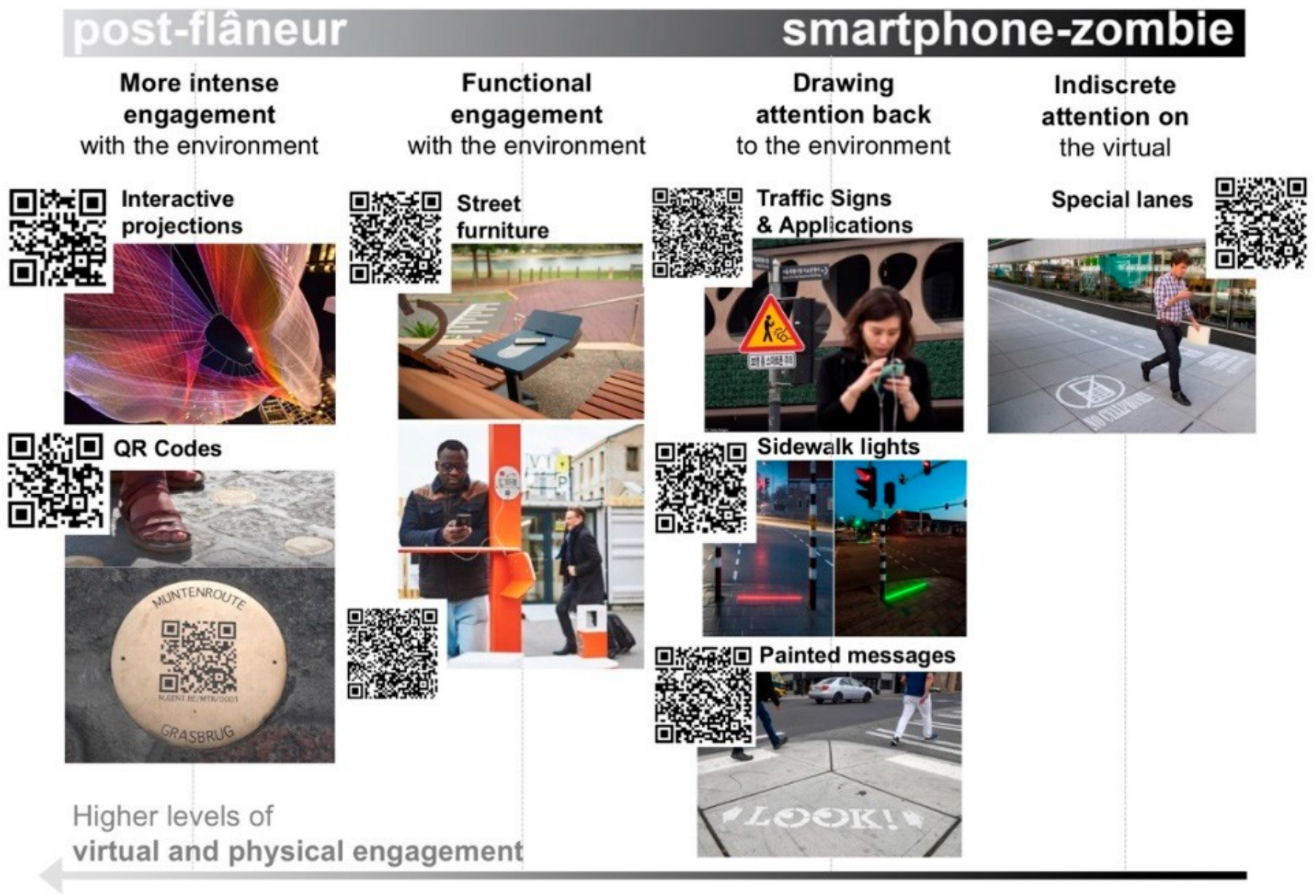

**Figure 2.** Existing public space design interventions addressing smartphone users [51–58] (original figure by the authors, images used can be reached through the references and QR codes).

Other interventions lie in the middle of the spectrum, and they aim to provide functional engagement with the environment. Street furniture for charging smartphones [55,56] can be listed among such design interventions that address smartphone users in public spaces. Other interventions involve higher levels of physical and virtual engagement. Some of them appear as art installations, such as interactive projections [57] that provide more intense engagement with the environment by projecting the virtual onto the real environment. These kinds of hybrid public space trials provoke loitering in the public space, thus targeting the post-flâneurs. Another popular example of such interventions is the use of QR codes in public spaces. An example of this is also seen in the area where the observations were carried out for the present study. As a part of the "Muntenroute" project [58] in Ghent city center, aluminum–bronze coins were placed on the pavement to represent a medieval trade route with drawings and descriptions accessed by QR codes found on the coins.

The interventions addressed above arose from the need to meet the new requirements of everyday life in public spaces transformed by mICTs and ubiquitous technologies. However, the transforming pedestrian movement patterns imply that the design of public spaces should be reconsidered by taking into account these new figures as well as altering behaviors, and a spatial design approach that goes beyond temporary additions to existing public spaces should be adopted.

## 3. Study Design

Korenmarkt Square—the most central and historical public space in Ghent, Belgium—was chosen as the case study area. This selection was based on the geolocation of social media use [59], the scale, and the movement patterns. A specific part of the public space was selected on account of its favorable scale for observation of the pedestrian flows, both in the forms of a street and a square (Figure 3). To observe pedestrians who use smartphones, we recorded 10-min videos of the Korenmarkt Square from the same spot for each day of the week and observed people who walked between the designated observation gates with a visible smartphone (Figure 3). The entire data-decoding process was performed manually by a single researcher through high-quality videos of the area that allow zooming without a significant data loss. Afterwards, a statistically significant sample of 153 smartphone users was cross-validated by an independent researcher. The results of this validation have been used to verify the reliability of the observation and codification methods, which are further discussed after Figure 3 in this section and under Section 4.

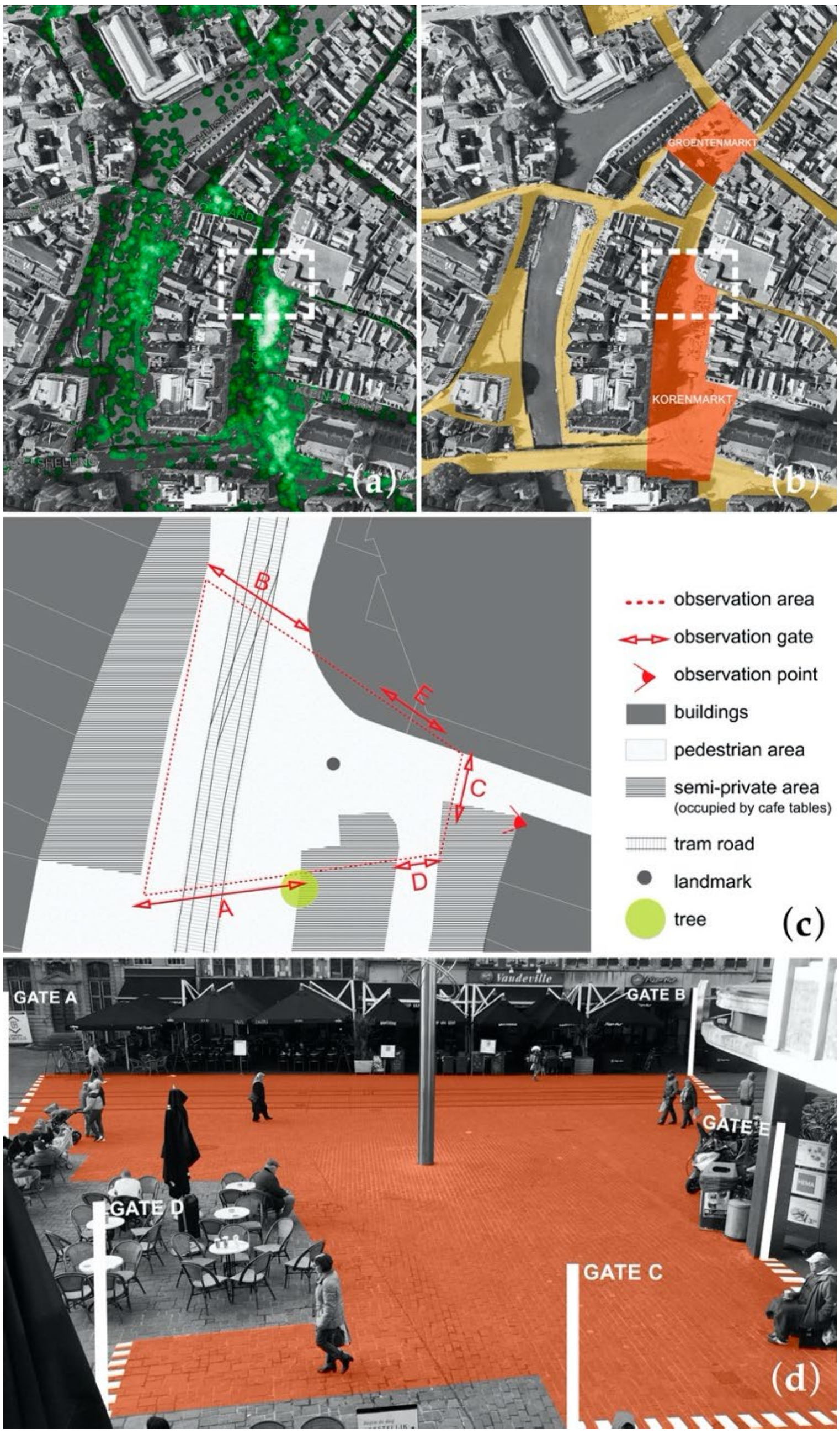

**Figure 3.** (**a**) Superposition of the six billion tweets map [59] on the Korenmarkt Square made by the authors, (**b**) pedestrian route network, public squares, and observation area (dashed white line), (**c**) map of the observation area with the boundaries of the camera perspective, and (**d**) the view from the observation point with the designated observation gates (original figure by the authors).

There were three main rules for observation: the person should (1) hold a smartphone visibly, (2) enter and exit from one of these gates within the determined time interval, and (3) be visible throughout the entire walk. If a smartphone user enters or exits from any other place apart from the observation gates (i.e., a cafe or restaurant) or a tram, or another pedestrian blocks our view of the user, they are excluded from the data.

Time intervals for the records were determined according to the maximum use of the area and the surrounding places with the help of popular times graphs derived from Google. The primary reason behind this was our intent to observe more pedestrians and, therefore, to detect more smartphone users. In addition, analyzing each day of the week allowed us to observe the area's less crowded times, since there was a sharp distinction between weekdays and weekends in terms of the pedestrian population. By limiting within the daytime and sunny days to enable a detailed observation, time intervals for each day were designated between 3–5 pm, and data representing one week were collected between 20 October and 14 December 2018.

Eventually, 455 smartphone users (278 female, 177 male) were detected. Previous studies that observed smartphone users in real-world environments accept that walking paths between 20–30 m are appropriate sightlines to observe walking [19] and stationary activities [30]. In the Korenmarkt example, the primary route is a straight line between observation gates A and B, and its length ranges from 22–32 m depending on the camera perspective. Therefore, 20 m could also be defined as the minimum route length to observe in this study. After the users with a walking path under 20 m (n = 104) and runners (n = 2) were excluded, we examined 350 smartphone users (198 female, 152 male) whose total length of walking path was 26.3 ± 4.1 m in average. The predicted gender and age group, number of companions, and their dominant mode(s) of smartphone usage were identified for each person. These smartphone-related activities were specified as follows: only holding (without using), only checking, listening, speaking, reading, typing, navigating, photo-taking, and video-recording. These activities can be observed from a distance since the usage of device itself is visible in the gestures and actions of walkers [15].

For temporal analysis, each person's total time between their entrance and exit was divided into individual seconds, and each second was coded according to their gaze directions by specifying whether it was a walking or a stationary activity. For this purpose, 24 different codes were created, and these were categorized under three different gaze types: gaze on the environment, gaze on the environment through the smartphone screen, and gaze on the smartphone screen (Table 1). An independent non-architect graduate researcher cross-validated a sample selected randomly from these 350 smartphone users. The sample size (n = 153) was determined using Cochran's Formula with a 90% confidence interval and 5 percent—plus or minus—precision. The cross-validations show that the disagreement levels were relatively low, with a 3.92% in smartphone activities, a 1.96% in age groups, a 1.31% in the number of companions, and a 0.65% in gaze directions/durations.

**Table 1.** Codes for different gaze types according to activities.

| | | |
|---|---|---|
| **GAZE 1** gaze on the environment | *walking* | (C1) gaze on the destination or a companion, (C2) showing directions, (C3) wandering gaze, (C4) photo and video hunting, (C5) speaking on the phone, (C6) speaking on the phone by holding in front of the face |
| | **stationary** | (C7) gaze on a companion or an object, (C8) showing directions, (C9) wandering gaze, (C10) photo and video hunting, (C11) speaking on the phone |
| **GAZE 2** gaze on the environment through the smartphone screen | *walking* | (C12) recording video, (C13) taking photo, (C14) taking selfie |
| | **stationary** | (C15) recording video, (C16) taking photo, (C17) taking selfie, (C18) posing for a photo |
| **GAZE 3** gaze on the smartphone screen | *walking* | (C19) gaze on the screen, (C20) sharing own screen, (C21) speaking on a video call |
| | **stationary** | (C22) gaze on the screen, (C23) sharing own screen, (C24) sharing someone's screen |

The walking routes and exact locations of stationary activities of smartphone users were mapped and joined with the gaze data in the Geographic Information System (GIS) to perform spatial analysis and calculate average walking speeds. The video recording was made by a high-resolution camera. The rates of the smartphone users were mapped in GIS manually by a single researcher following the methodology of William H. Whyte in his book *Social Life of Small Urban Spaces* [45]. Case-specific locations and traces on pavements visible in the video recording and aerial photos were mapped on the GIS basemap to correctly transfer walking routes and stationary locations by controlling their distance and proximity to these reference points.

After decoding and mapping, descriptive statistics, spatial analysis, and two-step cluster analysis (TCA)—which is a scalable cluster analysis method that uses an agglomerative hierarchical clustering method to group similar data by pre-clustering the data into many small sub-clusters and grouping the sub-clusters into a proper number of clusters automatically [60]—were used to analyze the data (Figure 4), the results of which will be summarized in the following sections of this paper. Statistical analyses were performed with SPSS (v25.0) for Mac OS, while spatial analysis and their visualizations were made in QGIS (v3.6) for Mac OS.

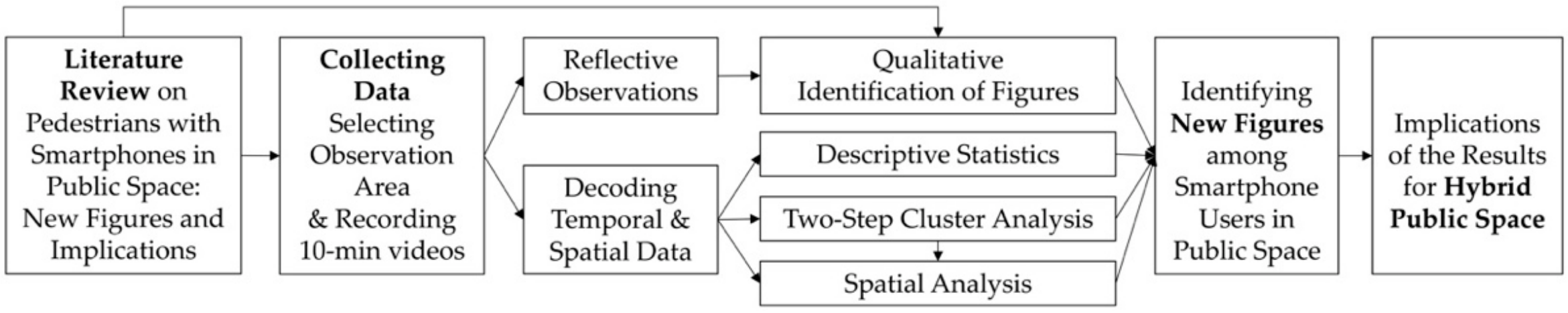

**Figure 4.** Research methodology.

## 4. Results

### 4.1. Descriptive Statistics

#### 4.1.1. Demographic Characteristics of the Passers-By and Smartphone Users

During the whole observation, 5809 pedestrians (3483 female, 2326 male) passed through the case area, and 455 of these (278 female, 177 male), which corresponds to 7.8% of the overall population in the area, were walking with a visible smartphone (Table 2). There was an evident difference between weekdays and weekends in terms of the number of smartphone users. On weekdays, when the observation area has a rather low pedestrian density, the percentage of smartphone users reaches up to 9.6%, whereas the percentage falls to 6.1% during the weekend.

**Table 2.** Number of all passers-by and smartphone users by predicted age groups and gender.

| | Number of Passers-By | | | Number of Smartphone Users | | | Percentage of Smartphone Users | | |
|---|---|---|---|---|---|---|---|---|---|
| | Female | Male | Total | Female | Male | Total | Female | Male | Total |
| Teenagers | 210 | 114 | 324 | 56 | 17 | 73 | 26.67% | 14.91% | 22.53% |
| Young adults | 435 | 284 | 719 | 113 | 64 | 177 | 25.98% | 22.54% | 24.62% |
| Adults | 2197 | 1453 | 3650 | 98 | 77 | 175 | 4.46% | 5.30% | 4.79% |
| Elderly | 641 | 475 | 1116 | 11 | 19 | 30 | 1.72% | 4.00% | 2.69% |
| Overall | 3483 | 2326 | 5809 | 278 | 177 | 455 | 7.98% | 7.61% | 7.83% |

When we compare the proportion of smartphone users to all passers-by according to age groups, teenagers and young adults show the highest rates with over 20% in consistence with the previous research [36]. These percentages get even higher among females. For instance, on weekdays, the percentage of female teenagers who use smartphones while walking is almost 30%.

#### 4.1.2. Smartphone Users

The following part of the paper discusses the results obtained by analyzing 350 (198 female, 152 male) of 455 smartphone users (smartphone users with a walking path under 20 m and runners were excluded). The findings reveal that the percentage of people who use a smartphone as they walk with others (56.5%) is higher than that of the people walking alone (43.5%). This rate is even higher among teenagers (74.5% with accompany versus 25.5% when alone). These results conflict with the findings of Hampton et al. [44] who suggest that people are more likely to use a mobile phone as they walk alone. This indicates the alteration in the degree of smartphone's penetration into our everyday lives in public spaces in due course.

*Smartphone activities.* The majority of the smartphone users observed in the study were engaged in only one smartphone activity, while only a few users (n = 42) were engaged in two smartphone activities. Analysis in this part of the study is performed for each individual smartphone activity, and, thus, smartphone users who were using two modes of smartphone activities were counted twice.

When it comes to different smartphone activities, we observed that reading (28.3%) and typing (16%) take the lead, followed by only checking (13.7%), speaking (12.6%), and navigating (12%), respectively. The high percentage of only holding (11.7%) shows that "phone-walkers" are one of the emergent modes of smartphone users in public spaces, which needs further attention, as suggested by Schaposnik and Unwin [30] and Fernández et al. [38]. Photo-taking (8.9%) follows only holding, and, lastly, listening (4.9%), and video-recording (4%) are the least used modes of the smartphone during walking and transient stationary activities in the public space. During the observations, only people who visibly used their smartphones for listening were taken into account, which explains the low percentage of listeners in the findings. The cross-validation results made by the second independent researcher indicate that the observation and codification methods employed were reliable. Cross-validation of 153 observed smartphone users showed low disagreement levels between two different researchers. Among the activities observed, detecting navigating smartphone users was the hardest, followed by the activities of typing and reading which were confused with each other in a limited number of cases. These disagreements and errors did not lead to statistically significant differences between the observation and codifications of the two independent researchers.

*Visual attention and movement.* 74.3% of smartphone users walked without stopping while others (25.7%) paused along their route. 84.8% of those who paused used their smartphones while stationary. Only 2.9% of smartphone users used their smartphones only when they stopped in the entire sample.

Results also show that 64.1% of smartphone users' gaze was on the environment (Gaze 1), while 35.9% gazed on and through the smartphone screen (Gaze 2 and Gaze 3) during the total duration of the walking and stationary activities. The percentages of visual attention during walking indicate similar results only with a lower percentage of gaze on the environment through the screen (Gaze 2) (1.1%). However, when we made an analysis with a focus on visual attention during stationary activities, we found that smartphone-based visual attention was higher (58.6%).

*Walking speed.* Walking speed is another essential variable to differentiate the figures among smartphone users in public spaces. The average walking speed for smartphone users was $1.09 \pm 0.29$ m/s. Due to the change in pedestrian density, the average walking speed varies on weekdays and weekends such that the total number of people counted on weekdays is almost equal to the number of people that passed through the same area on Saturday and Sunday (Figure 5). The highest average walking speed was observed on Monday as $1.24 \pm 0.27$ m/s, while the lowest occurred on Saturday at $0.95 \pm 0.27$ m/s. Regardless of the presence of a crowd that could affect the speed, the average speed of those using smartphones is lower than the average walking speed of pedestrians (1.4 m/s).

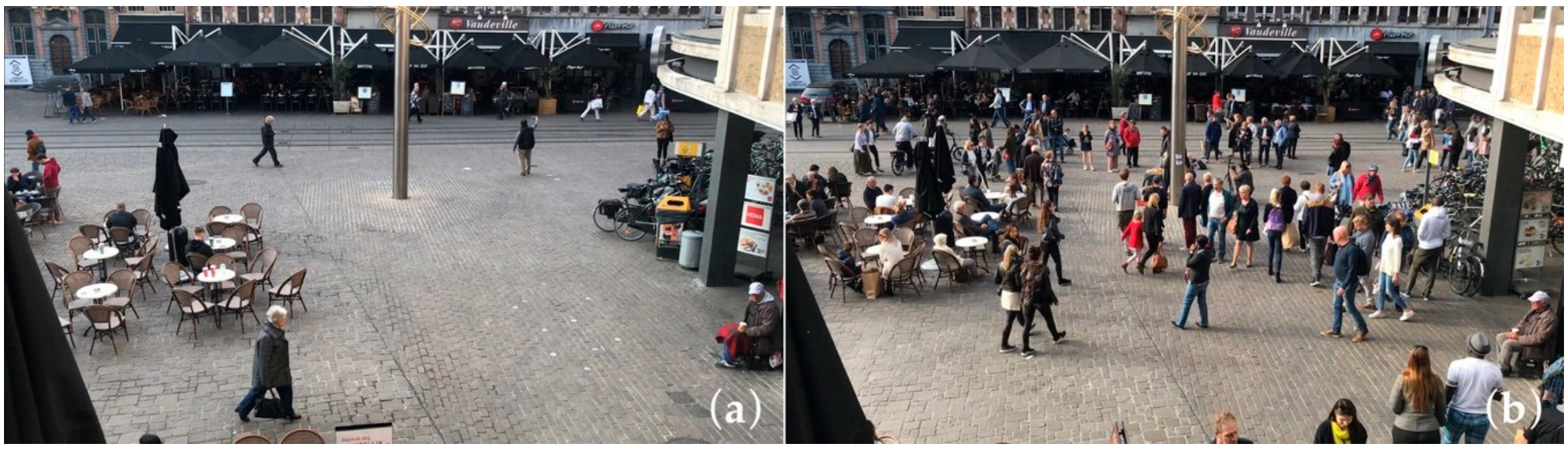

**Figure 5.** Comparison of the pedestrian density on Monday (**a**) and on Saturday (**b**).

*Wandering gaze.* We used the wandering gaze (calculated by using the codes of wandering gaze, photo/video hunting, and showing direction) to differentiate the active visual attention on the environment from a gaze on the destination or companion(s). Wandering gaze is especially essential to detect post-flâneurs. As Gros [47] (p. 210) mentioned, "The flâneur's body moves slowly, but his eyes dart about and his mind is gripped by a thousand things at once." Based on this, one can claim that post-flânerie also involves a high level of gaze wandering.

When we examined the mean percentages of visual attention while walking and the walking speed for each smartphone activity, an inverse correlation between speed and percentage of gaze on the smartphone screen was prevalent for the following activities: only holding, only checking, listening, reading, and typing. The activity of speaking goes as an exception to this rule. Although speaking does not take much visual attention, it can create a cognitive distraction [36], thus decreasing the walking speed. In addition to these, navigating, photo-taking, and video-recording deviate from other activities with their low percentages of gaze on the screen and low walking speeds. At this point, high percentages of wandering gaze associated with these activities provides a potential explanation for decreased speed (Figure 6).

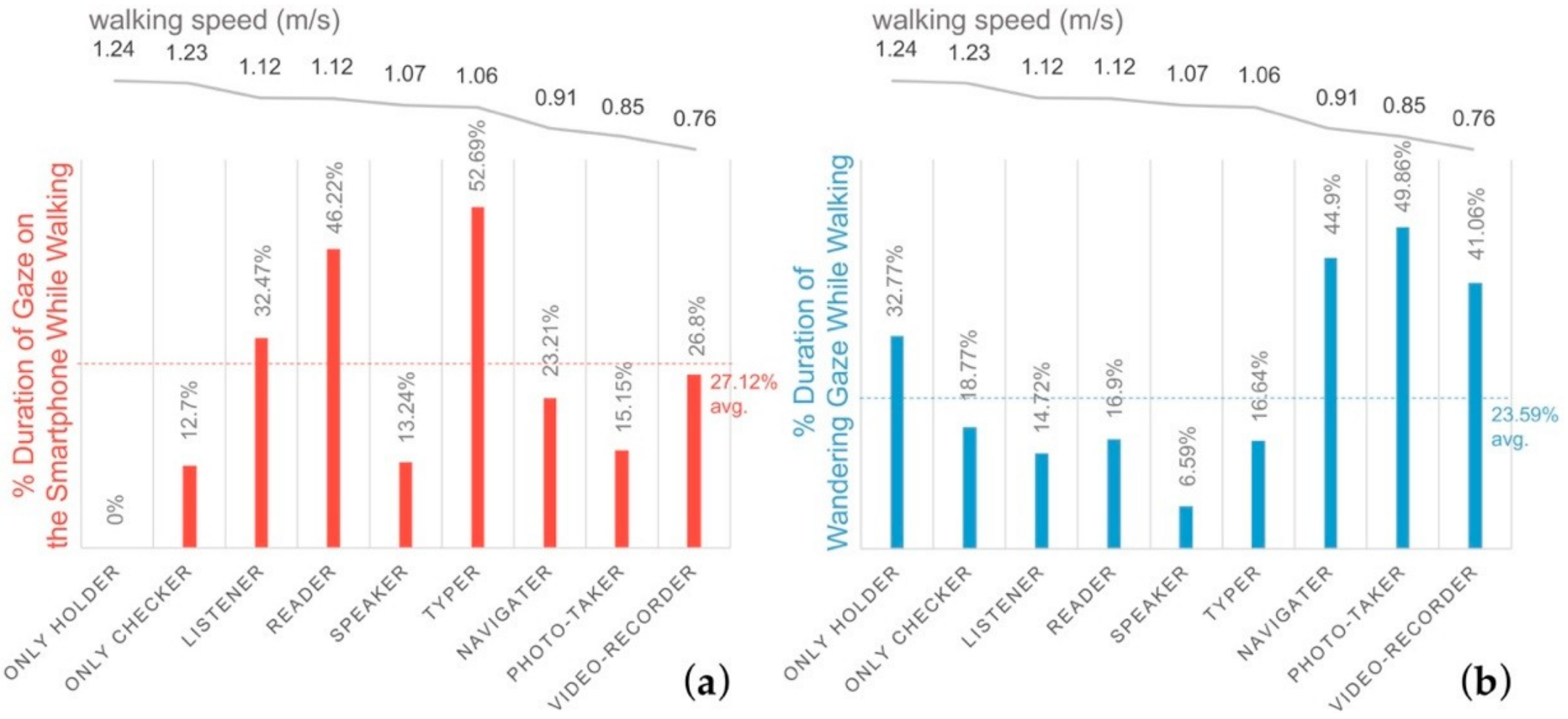

**Figure 6.** Mean percentages of (**a**) gaze on the smartphone screen and (**b**) wandering gaze while walking, and the mean walking speed for each smartphone activity.

In the following part of the paper, the nine smartphone activities we observed are grouped into four categories based on their similarities in gaze and walking behavior to facilitate further analysis. These groups are (1) listening and speaking, (2) reading and typing, (3) photo-taking, video-recording and navigating, and (4) only checking and only holding.

### *4.2. Detecting New Figures by Using Two-Step Cluster Analysis*

In this part of the study, we used TCA to detect the different figures in public space by clustering smartphone users according to their altering gaze types. We preferred utilizing TCA since it automatically determines the number of clusters with the best cluster quality option. For the purposes of this analysis, smartphone users (N = 350) were first divided into two groups made up of (1) people who walked without a pause (N = 257) and (2) people who stopped along their route (N = 93), and these groups were analyzed separately. The input variables for the TCA were gaze on/through the smartphone screen (%) and wandering gaze (%) during walking. For the TCA of the second group, we also included the same variables calculated for stationeries in the analysis. Variables of walking speed and smartphone activity type (categorized in four groups stated at the end of the former sub-section) were also included in the TCA as evaluation fields to see whether they are explanatory for the clusters. We used Euclidean distance coefficients for the variables and Akaike's Information Criterion (AIC) to determine the importance levels of variables in clustering. When the TCA was performed, the first group was divided into five clusters (av. silhouette = 0.65), while the second group was divided into four clusters (av. silhouette = 0.55) (Figures 7 and 8). The absolute distribution graphics in Figures 7 and 8 demonstrate the groups' distribution in general in terms of related variables. The background colors of the absolute distribution graphics indicate input (predictor) importance of the variables. Among the input variables, the percentages of wandering gaze (both when walking and stationary) have the highest importance level (importance = 1.00) in both groups, while the importance of the percentage of gaze on/through the screen also have significantly high importance ($0.75 \leq$ importance $\geq 0.94$). Evaluation fields of smartphone activities and walking speed have lower importance in explaining the clusters.

| | LABEL | | W1 | W2 | W3 | W4 | W5 |
|---|---|---|---|---|---|---|---|
| | Cluster Size | N | 34 | 33 | 55 | 104 | 31 |
| | | % all | 9.71 | 9.43 | 15.71 | 29.71 | 8.86 |
| INPUT | % Wandering Gaze (walking) | *M* | 87.16 | 2.60 | 4.98 | 1.26 | 45.11 |
| | | *SD* | 12.72 | 6.18 | 10.89 | 4.37 | 9.89 |
| | | Absolute Distributions | wandering gaze (w) | wandering gaze (w) | wandering gaze (w) | wandering gaze (w) | wandering gaze (w) |
| | % Gaze on/through the Screen (walking) | *M* | 6.64 | 82.39 | 49.30 | 7.70 | 27.58 |
| | | *SD* | 8.24 | 9.85 | 11.28 | 8.79 | 13.98 |
| | | Absolute Distributions | gaze on/through screen (w) | gaze on/through screen (w) | gaze on/through screen (w) | gaze on/through screen (w) | gaze on/through screen (w) |
| EVALUATION FIELD | Smartphone Activities | Most Frequent Category | #4 Only holding and checking (52.9 %) | #2 Typing and reading (90.9 %) | #2 Typing and reading (83.6 %) | #4 Only holding and checking (52.9 %) | #2 Typing and reading (45.2 %) |
| | | Absolute Distributions | smartphone activities | smartphone activities | smartphone activities | smartphone activities | smartphone activities |
| | Walking Speed | *M* | 1.08 | 1.20 | 1.17 | 1.24 | 1.09 |
| | | *SD* | 0.24 | 0.24 | 0.23 | 0.26 | 0.28 |
| | | Absolute Distributions | walking speed | walking speed | walking speed | walking speed | walking speed |

Input (Predictor) Importance ■1.0 ■0.8 ■0.6 ■0.4 ■0.2 □0.0

**Figure 7.** Results of centroids for the smartphone users who walked without stopping (n = 257). Absolute distribution graphics were generated with SPSS (v25.0).

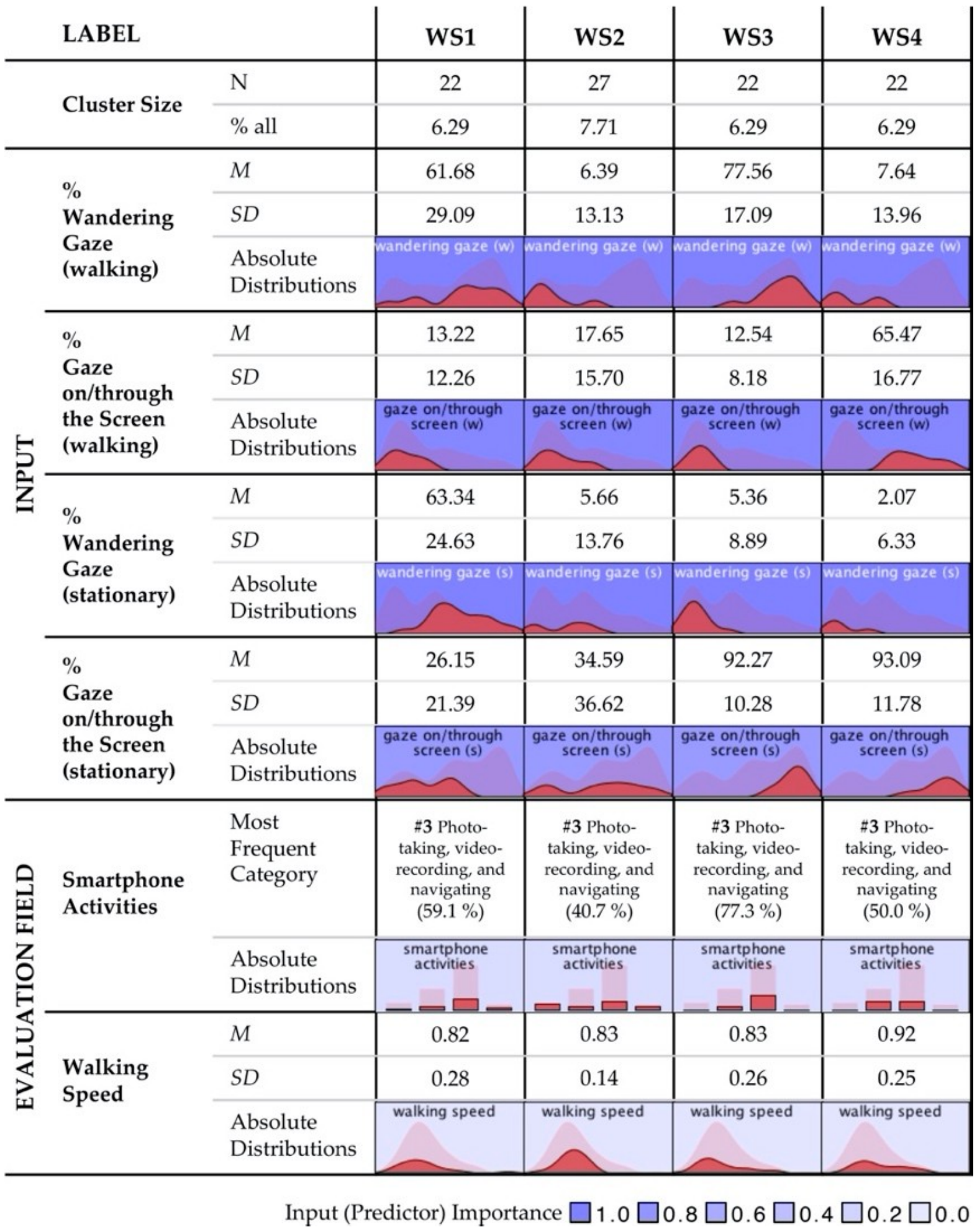

**Figure 8.** Results of centroids for the smartphone users who stopped along their routes (n = 93). Absolute distribution graphics were generated with SPSS (v25.0).

The TCA results for the smartphone users who walked without a pause (Figure 7) show that Cluster W4—which can be called "destination-oriented walkers"—constitutes approximately 30% of all smartphone users. They are the fastest group with the lowest percentages of wandering gaze and gaze on/through the smartphone screen. This cluster is mostly made up of only holders and only checkers (52.9%), and listeners and speakers (29.8%).

Cluster W1, with high percentages of wandering gaze, low percentages of screen-based visual attention, and low walking speed, fits best with the definition of "post-flâneur". With 52.9%, the majority of this group consists of only holders and only checkers, followed by navigators (26.5%). Cluster W5 follows post-flâneurs by spending half of their walking time with wandering gaze on the environment and walking slower than the average. Since their visual attention is divided between environment and screen while walking, they can be called "in-between walkers".

On the opposite end, Cluster W2 is the most appropriate to be referred to as "smartphone zombies" with their high percentages of gaze on/through the screen with an average of 82.3%. Almost all of the smartphone zombies (90.9%) were using their smartphones for typing and/or reading as they walk. They are followed by Cluster W3, which is made up of people who have low percentages of wandering gaze and spend almost half of their walking time with their gaze on/through the screen, mostly for reading or typing. They can be called "second-degree smartphone zombies".

The TCA results for the smartphone users who stopped during their walks (Figure 8) show that they can be grouped under four clusters according to their direction of gazes during their walks and stationaries. Cluster WS2 can be called "disinterested walkers" since they have low wandering gaze and screen-based visual attention whilst walking. However, the percentage of their gaze on the screen increases during their stationary time; 40.7% of them use their smartphones to take photos/record videos and navigate, implying that they probably use smartphone as a waiting activity.

Cluster WS4 is "immersed smartphone zombies" with a high percentages of gaze on/through the screen as they walk (*M* = 65.5%). This percentage skyrockets during their stationeries, with an average of 93.1%. This group has the highest walking speed among the smartphone users who stopped along their routes and consists of photo/video-takers, navigators, and typers/readers.

Cluster WS1 and WS3 suit for "stationary post-flâneurs" and "immersed post-flâneurs" respectively with their high percentage of wandering gaze on the environment while walking. These two clusters differ from each other on the basis of their gaze on/through the screen when they stopped. The former has a high average percentage of wandering gaze, while the latter has a high percentage of gaze on/through the smartphone screen during their stationary time. Both of these clusters are mostly comprised of photo-takers, video-recorders, and navigators.

Figure 9 summarizes and visualizes the clusters mentioned above by locating each smartphone user in a three-dimensional scatter graph of the percentages of wandering gaze, the percentage of gaze on/through the smartphone screen, and walking speed. In the following part of this section, we will focus on post-flâneurs and smartphone zombies to examine how they differentiate from each other and within themselves.

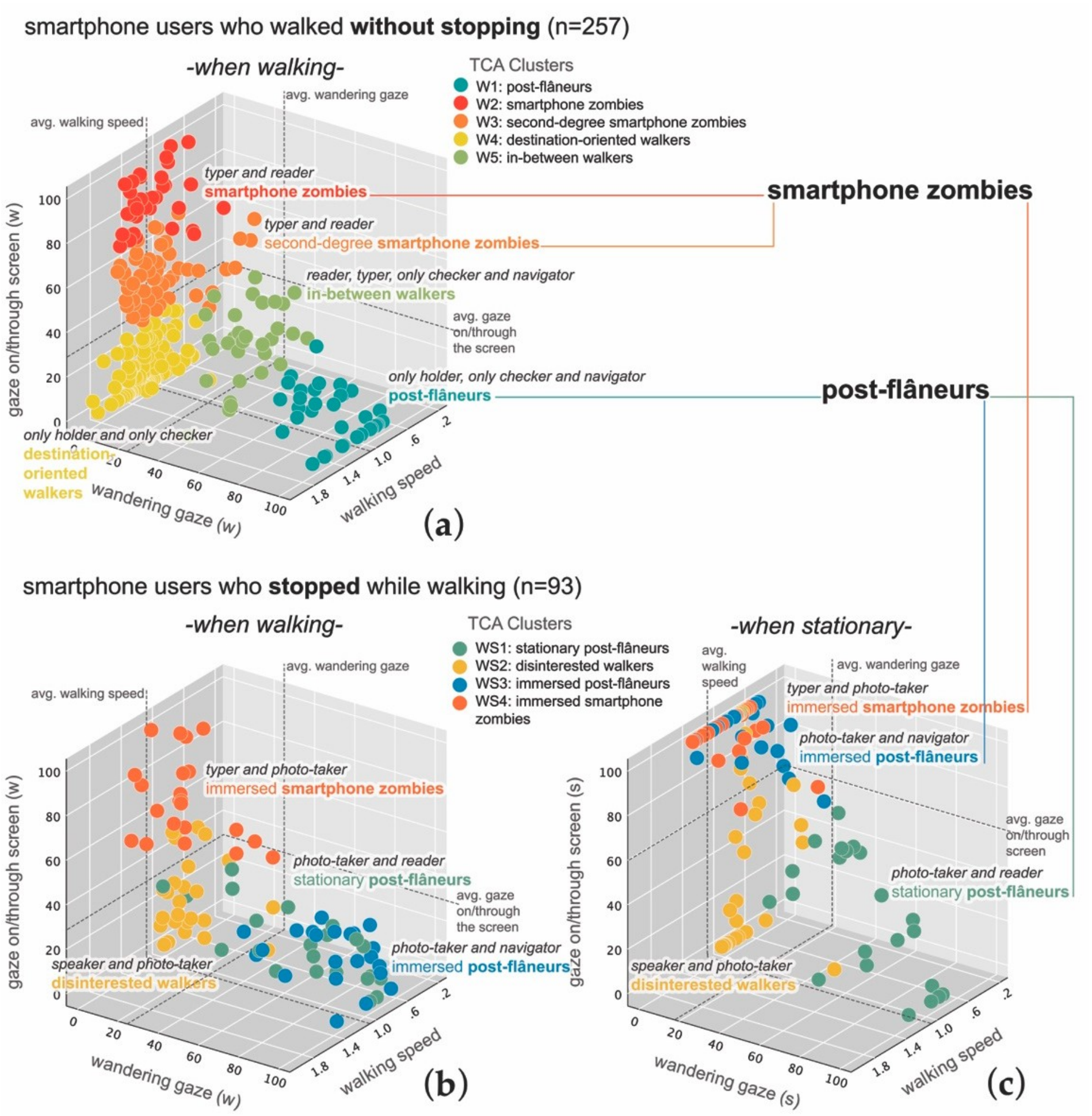

**Figure 9.** Three-dimensional scatter graphs of clusters generated by the two-step cluster analysis (TCA) within (**a**) smartphone users who walked without stopping and (**b**,**c**) who stopped while walking. The graphs represent the percentages of visual attention (**a**,**b**) when walking and (**c**) the stationary and walking speed of smartphone users. Scatter graphs were generated with SPSS (v25.0).

#### 4.2.1. Post-Flâneurs

Flânerie and post-flânerie involve aimlessly wandering along with high levels of gaze wanderings. Post-flânerie is unique in the sense that it also involves the simultaneous appropriation of smartphones. Using this behavior as a marker, we identified three groups of "post-flâneurs" in public spaces. The first group (W1) walks in the public space with intensive visual attention to the environment (more than 60%). As the concept of flânerie is always related to low walking speed (also known as "turtling"), post-flâneurs we observed also walk slowly (<1.4 m/s) while holding their smartphones without using or interacting with their smartphones only for checking and navigating. Post-flâneurs walks slower than destination-oriented walkers who are also only holders or checkers. We observed that they walk even slower than the smartphone zombies who concentrated on their screens as they walk.

The second group, called "immersed post-flâneurs" (WS3), exhibited similar behaviors while walking; however, they loiter more in public space and use their smartphones during stationary activities. They mostly consist of photo-takers and video-recorders; however, they not only take photos and record videos with their smartphones but also experience the public space through their screens. This indicates that, although they have the highest average lingering times in the public space (average of all stationary times), they do not spend this time only for engaging with the environment.

The third group, which can be referred to as "stationary post-flâneurs" (WS1), differs in this sense. These post-flâneurs are the most dispersed in terms of their percentages of wandering gaze. Their wandering gaze is higher than the average while walking, and it increases when they stop or vice versa. With immersed post-flâneurs, these two groups have the highest stationary times in the public space. These give a clear description of the characteristics and nature of post-flânerie behavior.

#### 4.2.2. Smartphone Zombies

This study proves the frequent presence of smartphone zombies in public spaces. In this study, we detected three kinds of "smartphone zombies". The first of these (W2) consists of typers and readers who walk in the public space with intense visual attention on their smartphone screen. Contrary to expectations, their average walking speed is higher than the average speed of all smartphone users observed in the study, and they have the second-highest walking speed following destination-oriented walkers. According to the results of the independent T-Test, there is a significant difference between the average walking speed of smartphone zombies ($M$ = 1.20, $SD$ = 0.24) and post-flâneurs ($M$ = 1.07, $SD$ = 0.24), $t(65)$ = 2.1, $p$ = 0.035, while there is no statistically significant difference between the walking speeds of smartphone zombies and destination-oriented walkers.

The second group is "second-degree smartphone zombies" whose percentages of gaze on/through the screen while walking is higher than the average (28%). This group also mainly consists of typers and readers. The last group, which can be named "immersed smartphone zombies" (WS4), includes the ones who stop and, in this sense, linger more in the public space. However, their visual attention is constantly on their smartphones during these stationary moments ($M$ = 93.09%). This group is slower than the first group of smartphone zombies; however, they have the highest walking speed among the pedestrians with smartphones who stopped while walking. Among smartphone users who made stop(s) while walking, this group makes the least number of stops along their routes. The immersed smartphone zombies can be further divided into two. Half of them use smartphones for typing and reading and remain stationary shorter than others (less than 10 s), which can picturize smartphone zombies in popular culture who stopped along their route as a result of intense attention on the screen. These findings indicate that "smartphone zombie" is more than a buzzword, and it is possible to identify at least three different types based on their behavioral patterns.

## 4.3. Spatial Analysis

### 4.3.1. Walking Activities

In scope of the spatial analysis, walking routes and standing locations of each smartphone user who walked more than 20 m (n = 350) were mapped and combined with the data-driven from the temporal analysis in QGIS. Afterward, the observation area was divided into a grid (0.5 × 0.5 m), and the average percentages of visual attention of the routes passing through each grid unit was calculated. The heatmaps below show the distribution of the walking routes of smartphone users in the observation area according to their percentage of gaze on the smartphone screen and percentage of wandering gaze while walking (Figure 10).

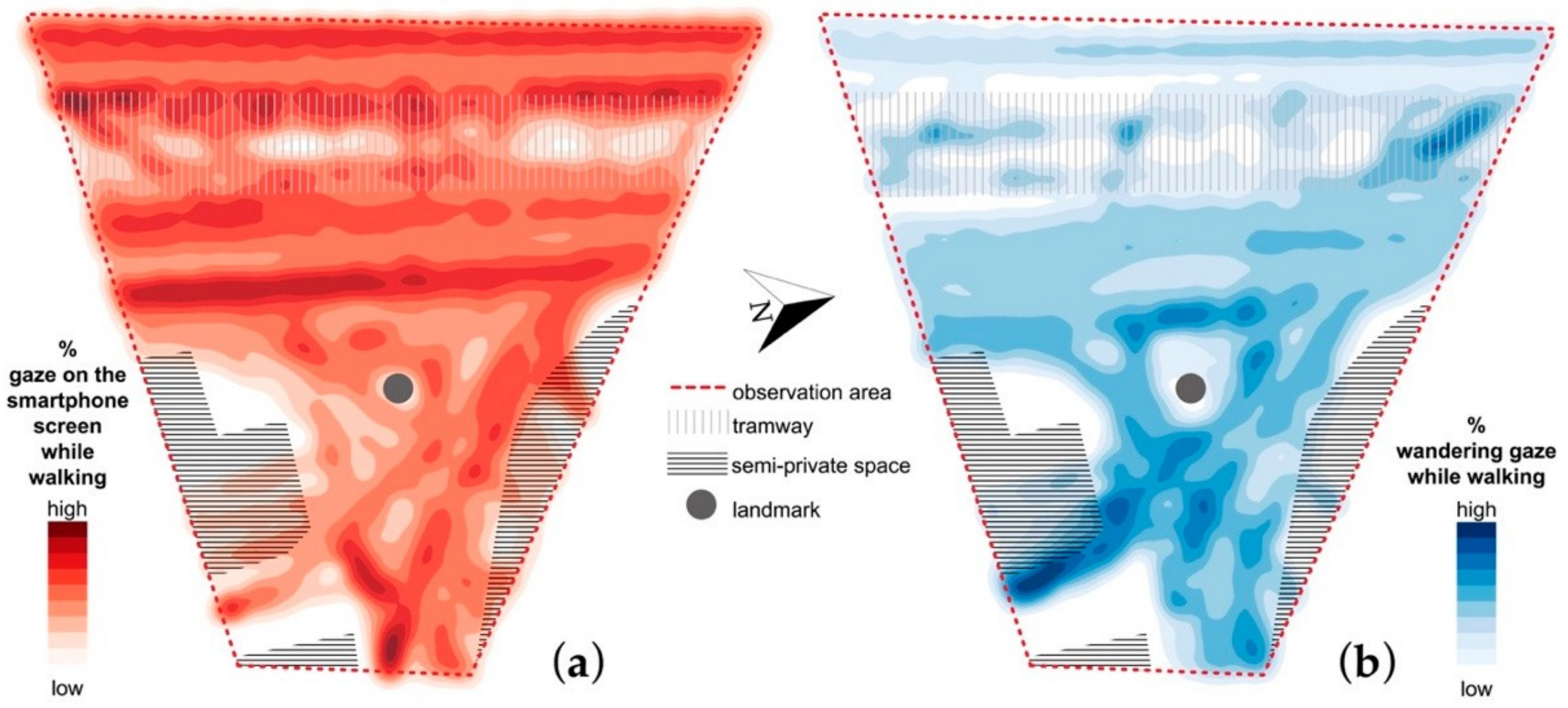

**Figure 10.** Heatmaps of the observation area that show the distribution of the routes of smartphone users according to (**a**) percentage of gaze on the smartphone screen, and (**b**) percentage of wandering gaze whilst walking. Maps were generated with the QGIS built-in heatmap renderer by using the quartic kernel shape function.

Figure 10 shows that the routes of people with high percentages of visual attention on the smartphone concentrate mainly on straight lines, whereas the routes of people who wander their gaze around the environment disperse in the open space surrounding the landmark in the middle of the observation area. It is clear that people who focus on smartphones as they walk prefer walking along the edges of the linear routes and the tramway—which actually does not constitute a physical obstacle—separates the movement of walkers. It is also worth mentioning that the routes of smartphone users with high smartphone-based visual attention are concentrated along the tramway, while the routes of people who cross the tramway have a rather high percentage of wandering gaze.

### 4.3.2. Stationary Activities

Stationary activities were visualized according to the total time spent in the public space while stationary, the duration of smartphone-based visual attention, and the duration of the wandering gaze (Figure 11). The analysis revealed that the smartphone users spent a majority of their stationary time with their gaze on their smartphones. This time, however, a higher amount of smartphone-based stationary activities are concentrated in the open space surrounding the landmark instead of straight lines as the case was in the smartphone-based visual attention in walking activities. The stationary wandering gaze displays a similar spatial distribution. However, the longest time spent while wandering gaze in the same spot in a stationary activity is 60 s, whereas the longest time spent for smartphone-based visual attention is 114 s.

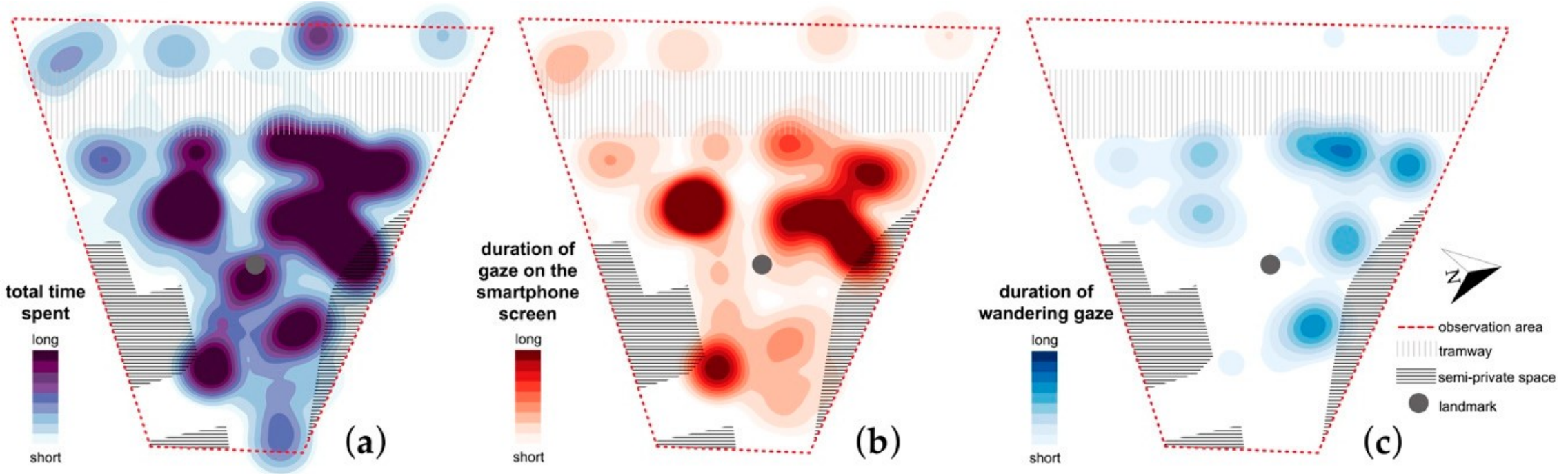

**Figure 11.** Heatmaps of stationary activities of smartphone users that show (**a**) total time spent, (**b**) duration of gaze on the smartphone screen, and (**c**) duration of wandering gaze. Maps were generated with the QGIS built-in heatmap renderer by using the quartic kernel shape function.

#### 4.3.3. Spatial Distribution of Movement of Post-Flâneurs and Smartphone Zombies

Pedestrian movement in public open spaces follows particular flows defined by boundaries, obstacles, and destinations. In the Korenmarkt example, semi-private spaces occupied by café tables, the entrance of a shopping center, and the tramway function as limiting elements for the movement in the space. Another factor that affects the movement in the observation area is the landmark (artwork) installed in the middle of the square. Apart from the triangle gap created around the landmark, the observation area consists of a dense network of flows. When we map the spatial distribution of smartphone zombies' movement, we see that they fit in this mostly-linear network of flows (Figure 12) and walk faster following destination-oriented walkers who mainly generate these flows.

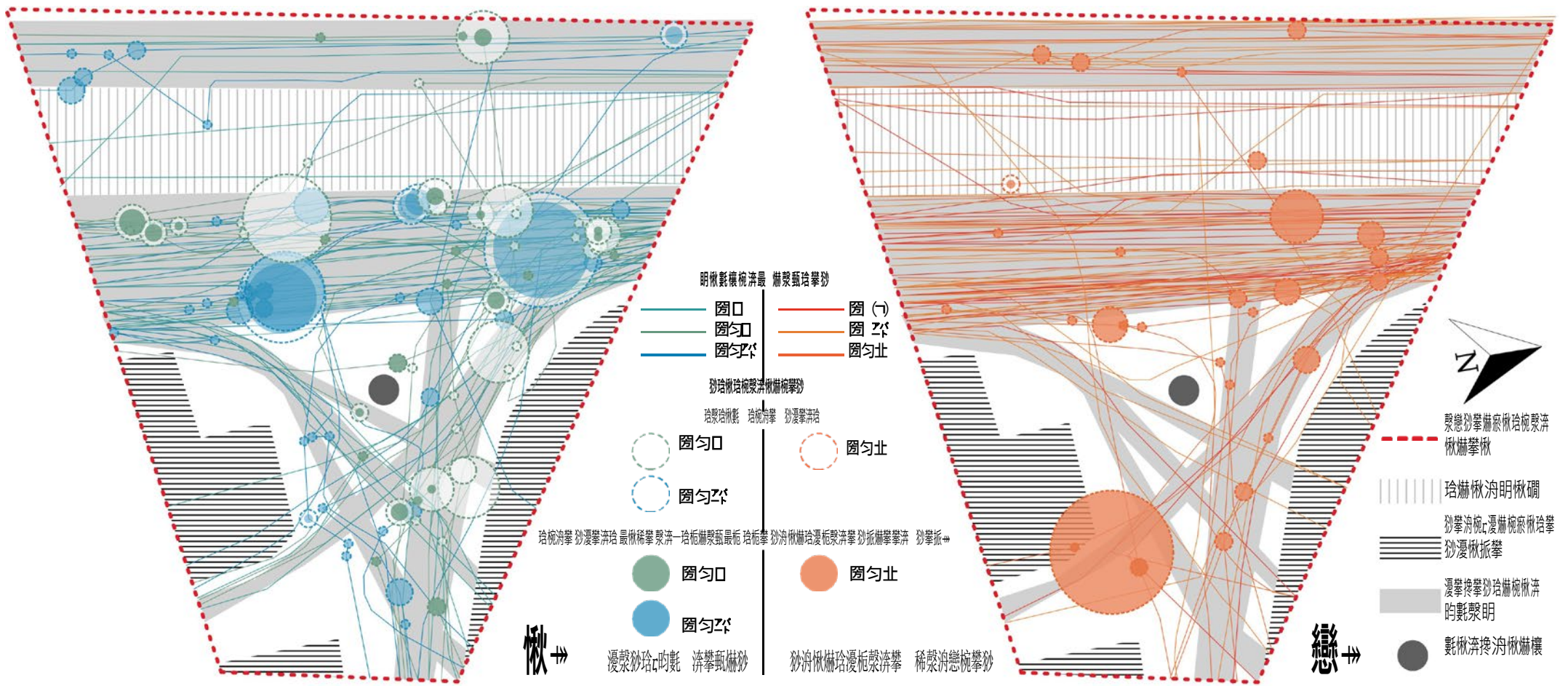

**Figure 12.** The distribution of walking paths and the duration of stationary activities of (**a**) post-flâneurs and (**b**) smartphone zombies. Stationary circles grow with increased waiting time. Maps were generated with QGIS.

Post-flâneurs, on the other hand, wander the most out of this network of flows, which is why they walk slower than the average walking speed of the pedestrians with smartphones observed in the study (Figure 12). Their movement spreads wider around the landmark. As opposed to the short pauses of smartphone zombies, post-flâneurs linger more in the public space. Unlike smartphone zombies, they (especially WS1) are not focused on the screen of their smartphones for a significant duration of their stationary time.

The comparisons between the spatial distribution of the movements of post-flâneurs and smartphone zombies highlight three major points:

1. Walking routes of post-flâneurs show a more organic pattern and concentrate around the landmark in the middle of the square. Smartphone zombies, on the other hand, walk in straight lines and concentrate on the rather linear part of the square.
2. Smartphone zombies move harmoniously with the flows in the public space, while post-flâneurs move independently from the mainstream.
3. Post-flâneurs stop more than smartphone zombies and present higher lingering times in the public space.

## 5. Discussion

The present study examined new modes of appropriation of public space by observing the altering visual attention and walking behavior of people who use smartphones in public space and identifying new figures and types of pedestrians. Results of this study indicate that pedestrians using a smartphone can be categorized by analyzing the altering direction of gaze (using the TCA method), and these figures can be associated with specific smartphone activities and walking speeds. The findings also reveal the nuanced characterization of novel spatial appropriations of the public space by these emergent new figures and how the spatial organizations of their movement differentiate from each other. In the following part of this section, we will address each of our research questions and arrive at conclusions based on our findings.

*RQ1. Identifying different figures among smartphone users in public space.* The findings of this study indicate that, using the presented method, pedestrians with smartphones in public spaces can be characterized as different figures according to the direction of their gazes and walking behaviors. (Figure 9, Table 3). In this study, two figures of the public space who position on the opposite ends of the virtual and physical space continuum (Figure 1) were identified with the help of TCA. Eventually, we detected three groups of post-flâneurs: (Pf1) post-flâneurs who walk slower than the average and wander their gaze around the environment, (Pf2) stationary post-flâneurs who focus more on the environment when they stop, and (Pf3) immersed post-flâneurs who focus on the screen when they are stationary. Smartphone zombies, on the other hand, are composed of three groups: (Sz1) smartphone zombies who walk faster than the average with intense attention on the smartphone screen, (Sz2) second-degree smartphone zombies who spend almost half of their walking time with their gaze on/through the screen, and (Sz3) immersed smartphone zombies who stop along their route and focus on their screen.

**Table 3.** Summary of the figures.

| Figure | N | % | Walking Speed | | Gaze Direction While Walking (%) | |
|---|---|---|---|---|---|---|
| | | | *M* | *SD* | Gaze on/ through the Screen | Wandering Gaze |
| Post-flâneur | 78 | 22.29 | 0.93 | 0.28 | 10.16 | 77.26 |
| Smartphone zombie | 110 | 31.43 | 1.13 | 0.26 | 62.46 | 4.80 |

These findings also highlight that the visual engagement of smartphone users with their environment varies proportionally with the estimated age group (Table 4). Decreasing age is accompanied by a decreasing wandering gaze and an increasing focus on the smartphone screen. A high percentage of teenagers observed in the public space were smartphone users, and a significant amount of them was defined as smartphone zombies (43.6%). Conversely, the percentage of smartphone users among elders was lower, and a significant amount of them was defined as post-flâneurs (53.6%). These findings implicating wide use amongst the younger population suggest that the future users of public spaces will probably wander through the virtual space more than ever while moving in the physical space. We should keep in mind that adults and elders we see in today's public spaces are still learning to walk with a smartphone. Young adults and especially teenagers, on the other hand, began their journey in the public space with a smartphone in their hands.

**Table 4.** Distribution of figures among the estimated age groups.

| Figure | Distribution of Figures among the Estimated Age Groups (%) | | | |
|---|---|---|---|---|
| | Teenagers | Young Adults | Adults | Elderly |
| Post-flâneur | 7.3 | 17.4 | 26.7 | 53.6 |
| Smartphone zombie | 43.6 | 34.8 | 28.1 | 7.1 |

*RQ2. The relation between smartphone users' visual attention and smartphone activities*. Different smartphone activities lead to different effects on visual attention. Previous studies revealed that the most common activities, such as reading and typing, proved to be the most distractive modes of using smartphones while walking [12,13,17,19,23,24,27,29]. Our findings reinforce this finding, since 64.9% of typers and readers we observed in this study were defined as smartphone zombies (Table 5). The findings also showed that not all smartphone activities isolate people from their surroundings. Smartphone activities such as navigation and photo-taking required active visual attention on the environment and revealed most of the post-flâneurs (48.1%) in this study (Table 5).

**Table 5.** Distribution of figures among smartphone activities.

| Figure | Distribution of Figures among Smartphone Activities (%) | | | |
|---|---|---|---|---|
| | Listening, Speaking | Typing, Reading | Photo-Taking, Video-Recording, Navigating | Only Holding, Only Checking |
| Post-flâneur | 6.5 | 11.2 | 48.1 | 23.6 |
| Smartphone zombie | 8.7 | 64.9 | 23.5 | 0 |

*RQ3. The relation between smartphone users' walking speed and movement in the public space and smartphone activities.* Previous studies found that smartphone distraction decelerates walking speed [12,14,32,34]; however, the extent of smartphone's influence over walking speed may depend on the intensity of pedestrian traffic, demographics, and the time of day [32]. By examining a central public space, this study found that people with smartphones walk slower than the average walking speed, and their speed decrease is inversely proportional to an increased focus on the smartphone screen, in parallel to these studies (Figure 6). Nevertheless, although walking slowly and being unresponsive to the surroundings causes pedestrians with smartphones to be called smartphone zombies, a significant number of those detected in this study were walking faster than the average speed of smartphone users (Figures 7–9, and Table 3). Despite these observations, the smartphone zombie epithet is still appropriate. As the fictional figure of "zombie" is present physically in our world while belonging to another world, a smartphone zombie has to be bodily present and walk in the physical space, while belonging to and wandering through the virtual space simultaneously.

Contrary to smartphone zombies, smartphone users such as navigators, photo-takers, and video-recorders present the lowest walking speed among the users of other smartphone activities, even though they have low percentages of smartphone-based visual attention (Figure 6). At this point, the wandering gaze and stationaries appear to provide an explanation for this observation. Smartphone-based visual attention decreases the walking speed; however, the wandering gaze may cause more decline in walking speed in comparison to smartphone-based visual attention.

*RQ4. Implications for hybrid public space and its design.* Recognizing these new modes of appropriation of public space raises questions on their design implications and possible approaches addressing the emergent "hybrid public space." As mentioned at the beginning of the study, mICTs have created personal hybrid spaces strolling around the physical public space, and currently, the most used interface for those hybrid spaces are smartphones. Today, public space itself can no longer be perceived regardless of the technology it contains and should be redefined and redesigned as a hybrid public space. In this process, mobile technologies should be included in and direct the design of the hybrid public space of the 21st century as a new dimension in the same way vehicles had radically changed urban design and planning at the beginning of the 20th century. In this sense, the hybrid public space should be handled and designed not only as a physical construct but as a hybrid construct in which the physical public space itself can turn into information technology and an interface. This way, instead of experiencing the virtual space through individual screens, we can create more collective modes of a hybrid experience and improve the equality aspect of the public space by improving the interaction not only between the virtual and the physical but also between people-environment-technologies. Furthermore, the hybrid public space of tomorrow can be imagined to provide meaningful interactions between the occupants of different public spaces. There have already been some examples such as "urban machines" [9], which can give ideas and inspiration about the implementation of the hybrid public space.

Today, public space design interventions addressing smartphone users mostly appear to be additions to the public space to solve the problems that arise from the high visual attention on the smartphone screen (Section 2, Figure 2). Apart from this, we also see examples of legal actions. Starting with banning texting while walking on crossings in Honolulu in 2017 [61], these legal actions reached another dimension in 2020 with the city of Yamato that banned using smartphones while walking in public space [62]. Time will tell how effective these will be. However, the emergent figures and new walking behaviors in public space revealed that we have to consider the public space's design beyond temporary additions to the existing public space or bans. Recognition of these figures and their projections in the urban space will be the first step to achieve this and constitutes this study's primary motivation.

This study reveals that the spatial organization of a post-flâneur's movement dissociates from a smartphone zombie also in the physical space. The extent of a post-flâneur's and smartphone zombie's wandering in the virtual space differentiate from each other. It means that the physical configuration of public space itself directly affects the orientation of visual attention, which underlies that city planners, urban designers, and all actors who play a role in designing public space should focus on creating possibilities for engagement with and wandering through the public space instead of developing obstacles. Today's public space still harbors obstacles, such as rapid flows and invasion of private spaces, even when it can partially escape from the hegemony of vehicles in the developed world.

The findings of this study also show that people whose gaze wanders on the environment wander not only with their eyes but also with their bodies. Thus, they not only use wider parts of the public space for stationary activities but also need these spaces for wandering their gaze whilst walking. This means that we need to design spaces for wandering—in other words, for the sole experience of walking—more than just designing for purposive walking [63]. To enhance engagement with public spaces, we need to produce spaces that allow more wandering and lingering and invite post-flâneurs. For this purpose, it is essential to design new and/or protect and expand the existing narrowed wandering spaces in public space, allow pedestrians to exist out of the rapid flows, and question the agencies of landmarks that widen the range of motion outside these flows.

Public space's hybridity is also critical to becoming a part of wandering and lingering. The findings of this study show that smartphone activities affect lingering time in public spaces and provide us a new position towards the debate between two discourses: "Cellphones, in this case, smartphones, depart us from the public space" by focusing attention to—mostly personal—virtual space [6,64] versus "they do not disconnect the society" since they make us spend more time in public space by increasing the lingering time [44]. In some cases, smartphones increase lingering in public space, as suggested. However, the present study shows that if a person with a smartphone stops in the public space while walking, they are more likely to use the smartphone and spend most of their stationary time on-screen. This implies that the phenomenon of private space interference in public space through smartphones deserves further attention.

The spatial distribution of movement of smartphone users also revealed that people who focus on the smartphone screen mostly walk in straight lines, similar to destination-oriented walkers. For instance, except for a few outliers, smartphone zombies observed in this study show up in the straight routes along which they can move without paying any attention to their surroundings. The comparison of weekdays and weekends also supports this by implying that the possibility of using a smartphone while walking decreases in dense public spaces. This can mean that, contrary to what Yoshiki et al. [19] suggest, smartphone users do not adjust their level of visual confirmation of the surrounding as they walk regardless of changes in pedestrian density, and in general, they do not prefer to look at their smartphones while walking in dense public spaces where the routes to follow are complicated. This is in conformity with the findings of a previous study [32]. If we consider that smartphone zombies prefer to walk in spaces that accommodate more linear pedestrian movements, breaking the linearity of their flows can be a solution to draw their visual attention back to the physical space, especially in the danger zones in public spaces when visual attention is vital. If the opposite impact is desired, space's linearity will be more directive for smartphone zombies than a phone lane drawn on the ground [54].

## 6. Conclusions

This study aimed to understand the visual engagement of smartphone users with the physical public space by examining their altering gaze and movement and identifying new figures among them. To achieve this, we conducted an in-depth literature review on pedestrians with smartphones, emerging new figures, and interventions in public spaces, identifying key theoretical concepts and hypotheses surrounding the subject matter. Building on this review, we carried out a natural observational study in which we observed a public space in the city center of Ghent, Belgium, for seven days in 10-min time intervals and analyzed 350 smartphone users among the passers-by. Consequently, we detected new figures among the smartphone users by using the TCA method, addressed them within the conceptual framework, and explored the answers to our research questions.

The knowledge contribution of this study is threefold. First, a major part of the literature on walking with a smartphone in public space approaches the topic from a safety perspective and examines smartphone users as "distracted walkers". However, this study shows that, when it comes to engagement with public space, a smartphone user is more than just a distracted walker. Second, this study made a methodological contribution that can be used in identifying new figures in public spaces through systematic video analysis. Third, observation-based identification of the new figures and recognizing how they appropriate public space paves the way to develop new approaches that address the emergent "hybrid public space".

At the extend of this study, we detected several figures among smartphone users and focused on the emergent figures of post-flâneurs and smartphone zombies. According to TCA results and our detailed observations, post-flâneurs walk slower than the average and typically wander their gaze around the environment while strolling in the public space. Some of them stop during their routes and either continue to interact with the environment visually or use their stationary times to focus on their smartphones, which mediate capturing/recording and navigating the environment. On the other hand, smartphone zombies walk with intense attention on the smartphone screen, usually to read and type something. Some of these smartphone zombies stop along their route and focus on their screen even more. An unexpected result was that smartphone zombies were walking faster than the average speed of smartphone users, and they were statistically faster than post-flâneurs. Spatial analysis shows that the fundamental reason behind this was that they were walking in harmony with the flows in public space, unlike post-flâneurs who wander out of these flows from time to time.

Developing a better understanding of these new behaviors in today's public spaces and their new modes of spatial appropriation gives us a chance to discuss and imagine the design of the hybrid public space of tomorrow, which is a "collectively inhabited urban space" that reconstructs the traditional interaction between the human and its physical and social environment [9].

During this study, only 7.8% of the observed pedestrians were using a smartphone (Table 2), and 31.4% of those we examine were identified as smartphone zombies (Table 3). This number may seem like a small proportion, and the need for specialized spatial interventions for smartphone zombies can be questionable. However, the increasing use of smartphones in everyday life and its reflections in public space breeds the need to consider its spatial dimension. This study also encourages this idea by underlining that "space is [still] the machine!" [65] in explaining and shaping the spatial configuration of pedestrians with smartphones in public space. This finding indicates that there can be more effective tools to organize and steer smartphone usage in public space beyond temporary additions or bans. From this perspective, we suggest that the hybrid public space should be realized by designing spaces for wandering, which creates possibilities for engagement with the public space both physically and visually, taking into account the public space's hybridity to become a part of the stationary time of the smartphone user, and last but not least, breaking the linearity of the pedestrian flows in the danger zones.

*Limitations and Future Directions.* In this study, we only focused on engagement with the public space through gaze and movement, and this proves to be the main limitation of the study. Further studies are necessary to provide a deeper perspective by focusing on the mutual relation between human–smartphone–environment and to enhance the research by examining altering audial and cognitional attention and engagement with the environment from the egocentric perspective of smartphone users.

Another shortcoming of the study is the lack of comparison between smartphone users and non-users. The primary reason behind this is the fact that the entire data-decoding process was performed manually, which was time-consuming. Automated data-decoding and analysis is developing technology and has come a long way since the MIT studies were held in the late 1990s [66]. It can be utilized to automatically detect smartphone users in a similar fashion to what Alsaleh et al. [12] did in their research. However, decoding data manually becomes essential to ensure asking the right questions and to understand and define what to look for. For instance, codes for gaze types in this study had taken their final form through the manual decoding process and reflective observations in the case area. Thus, further studies can make use of the method and descriptive variables defined for the figures of public space in this study and combine these with automatic data-decoding and analysis processes, and help with training machine learning algorithms. For this task, Dubba et al.'s [67] study provides a well-defined "supervised" video analysis framework using artificial intelligence. Learning from human understandable event models and employing these models to recognize events in videos recorded in our study can be supervised based on the analysis results and the conceptual framework presented in this study. Future research in this direction can help larger datasets to be studied and reveal novel spatial and temporal interval relations combined with pedestrian activities and experiences.

This study can be taken further to following Bhatt and Shultz's [68] transdisciplinary research on new generation Cognitive CAAD systems, linking architecture, its conception, and pedestrian experience. This framework provides solid methods for evaluating spatial design scenarios combining architecture and engineering, cognitive science, spatial cognition and computation, and empirical methods based on geo-information. Using this framework, the observations of the public space can be enriched as a novel design tool enabling investigations on continuity, visibility of landmarks, and circular and hierarchical spatial organization.

The nature of public spaces is changing with the advances in mobile technologies, and at the moment, smartphones with location-based applications are one of the major actors in this alteration. Slowly-evolving public spaces of the past have become more dynamic with invisible social interactions. They have already and will increasingly become more responsive, adaptive, multi-functioning, and flexible constructs that can adapt to the advances in technologies. To understand this change, there is a need to observe the new figures that foreshadow this alteration and spatial distribution of their movement in the public space using geo-located data, information, and spatial analysis methods. The rapid penetration of smartphones and location-based applications into public life and its effects on human behavior and public space shows that there will be other challenges, such as privatization and commercialization of the experience. However, by understanding these altering experiences and new figures, we can rethink the design of the hybrid public space, improve the relationship between humans, technologies, and the urban environment, and overcome these challenges.

**Author Contributions:** Conceptualization, Gorsev Argin, Burak Pak, and Handan Turkoglu; Methodology, Gorsev Argin, Burak Pak, and Handan Turkoglu; Data Collection and Curation, Gorsev Argin; Analysis, Gorsev Argin and Burak Pak; Writing—Original Draft Preparation, Gorsev Argin, Burak Pak, and Handan Turkoglu; Writing—Review & Editing, Gorsev Argin, Burak Pak, and Handan Turkoglu; Visualization, Gorsev Argin; Supervision, Burak Pak. All authors have read and agreed to the published version of the manuscript.

**Funding:** This research was funded by Research Fund of the Istanbul Technical University, grant number 39666. The processing fee for this journal paper is co-founded by KU Leuven, Sint-Lucas Brussels and Ghent Campuses.

**Acknowledgments:** This paper is produced as a part of a Ph.D. dissertation by the first author supervised by the second and the third authors. The authors would like to thank Baris Uz for his assistance in cross-validation of the decoded data, Belgin Gumru for her comments on the manuscript, and, last but not least, the anonymous referees for their useful suggestions, which helped to improve the quality of this manuscript.

**Conflicts of Interest:** The authors declare no conflict of interest. The funders had no role in the design of the study; in the collection, analyses, or interpretation of data; in the writing of the manuscript, or in the decision to publish the results.